\documentclass[12pt]{article}
  \usepackage{amsfonts}
\begin{document}
\begin{flushright}
IFIC-07-78, \quad FTUV-07-1112, \\ arXiv:0712.1826[hep-th]  \\
December 11, 2007
\end{flushright}

\vspace{1.0cm}

\begin{center}
{\Large\bf On covariant quantization of M0-brane. \\ Spinor moving frame,
 pure spinor formalism \\ and  hidden symmetries of D=11
supergravity} \\
\vskip 1.0cm  {\bf Igor A. Bandos}
 \\
\vskip 0.5cm { \it Departamento de F\'{\i}sica Te\'orica, Universidad de Valencia and
IFIC (CSIC-UVEG), 46100-Burjassot (Valencia), Spain
\\
Institute for Theoretical Physics, NSC Kharkov Institute of Physics  and Technology,
UA61108, Kharkov, Ukraine}

\vskip 0.5cm

Abstract
\end{center}
{\small The covariant quantization of massless D=11 superparticle (M0--brane) in its
twistor-like Lorentz harmonic formulation is used to clarify the origin and some
properties of the Berkovits pure spinor approach to quantum superstring and to search
for hidden symmetries of D=11 supergravity. In the twistor like Lorentz harmonic
formulation, the SO(16) symmetry is seen already at the classical level. The
quantization produces the  linearized supergravity multiplet as
$\mathbf{128+\widetilde{128}=256}$ component
 Majorana spinor of $SO(16)$ and
also shows an indirect argument in favor of the possible $E_8$ symmetry. }


\subsection*{1. Introduction}


In this contribution we briefly review the results of \cite{IB07PLB,IB07,BdAS2006} on
the covariant quantization of the M0--brane (D=11 massless superparticle)
\cite{B+T=D-branes,Green+99} in its twistor-like Lorentz harmonic or spinor moving
frame formulation \cite{BL98',BdAS2006}. We show how the covariant BRST quantization of
this model \cite{IB07PLB,IB07} explains origin and some properties of the Berkovits
pure spinor approach \cite{NB-pure} (see also {\it e.g.}  \cite{MT+pure,GrAV04} and
refs. therein), in the frame of which a significant progress in covariant loop
calculations has been reached \cite{NBloops}. Then we discuss how the covariant
quantization of physical degrees of freedom \cite{BdAS2006,IB07} shows the tails of
possible hidden symmetries of the eleven dimensional supergravity.

\subsection*{2. Brink--Schwarz superparticle and its $\kappa$--symmetry }

The Brink-Schwarz massless superparticle action, ($\Pi^m := dx^m - id\theta \Gamma^m
\theta = d\tau \hat{\Pi}^{m}_\tau$)
\begin{eqnarray}\label{11DSSP-1st}
S_{BS}  & =   \int_{W^1} \left(P_{{m}} {\Pi}^{m} -  {1\over 2}d\tau \; e\; P_{{m}}
P^{{m}} \right)\; ,   \qquad
\end{eqnarray}
possesses the local fermionic $\kappa$-symmetry \cite{dA+L82}, \cite{W.S.83}:
\begin{eqnarray}\label{kappa-r}
\delta_\kappa x^m = i \delta_\kappa \theta^\alpha \Gamma^m_{\alpha\beta}\theta^\beta\;
, \qquad \delta_\kappa \theta^\alpha = \tilde{P}\!\!\!/^{\alpha\beta} \kappa_\beta \; ,
\qquad \delta_\kappa P^m=0 . \qquad
\end{eqnarray}
This is necessary to provide that the ground state of the model preserves a part (1/2)
of supersymmetry and, thus, is a stable state - called BPS state (as they saturate a
Bogomol'nyi--Prasad--Sommerfield  bound). However this symmetry happens to be
infinitely reducible. Indeed, the $\kappa$-symmetry parameter of the form
$\kappa^{\!\!\!\!^{0}}_\beta=
\tilde{P}\!\!\!/^{\beta\gamma}\kappa^{\!\!\!\!^{1}}_\gamma$  does not produce any
transformations of $x^m(\tau)$ and $\theta(\tau)$. First one observes that
$\tilde{P}\!\!\!/^{\alpha\beta} \kappa^{\!\!\!\!^{0}}_\beta=
\tilde{P}\!\!\!/^{\alpha\beta}\tilde{P}\!\!\!/^{\beta\gamma}\kappa^{\!\!\!\!^{1}}_\gamma=
{P}_mP^m \kappa^{\!\!\!\!^{1}}_\alpha$ which vanishes due to the mass shell constraint
\begin{eqnarray}
\label{PmPm=0}  P_mP^m=0 \;
\end{eqnarray}
which appears as equations of motion for the auxiliary einbein field $e(\tau)$ in
(\ref{11DSSP-1st}). In the terminology of \cite{BV83} this is characterized by stating
that the symmetry ($\kappa$-symmetry in our case) has the {\it null vector}
($\kappa^{\!\!\!\!^{0}}{}^\alpha=
\tilde{P}\!\!\!/^{\alpha\beta}\kappa^{\!\!\!\!^{1}}_\beta$). The symmetry which has a
null vector is called reducible. Provided this null vector did not have its own null
vector, the reducibility would be  of the first rank and the effective number of
symmetry parameters would be equal to the number of manifest parameters minus the
number of null vector. If the null vector had its own null vector, but the latter did
not have the null vector of its own, the reducibility of the symmetry would be  of the
second rank and the effective number of symmetry parameters would be  calculated as the
manifest number of parameters menus number of null-vectors plus number of null-vectors
for null vectors, and so on.

For the $\kappa$--symmetry, one notices that the null--vector
$\kappa^{\!\!\!\!^{0}}{}^{\alpha}=
\tilde{P}\!\!\!/^{\alpha\beta}\kappa^{\!\!\!\!^{1}}_\beta$ has its own null--vector
$\kappa^{\!\!\!\!^{1}}_\alpha= {P}\!\!\!/_{\alpha\beta}\kappa^{\!\!\!\!^{2}}{}^\beta$
which has null--vector $\kappa^{\!\!\!\!^{2}}{}^{\alpha}=
\tilde{P}\!\!\!/^{\alpha\beta}\kappa^{\!\!\!\!^{3}}_\beta$ {\it etc.}. This chain is
infinite as $j+2$ null--vector coincides with the $j$-th one. Furthermore, all the
(reducible) null-vectors have also the same dimension equal to the number of component
$n$ of the minimal spinor representation ($\alpha,\beta=1,...,n$; $\; n=32$ for $D=11$,
$n=16$ for $D=10$, {\it etc.}). Then the effective number of supersymmetry parameter is
defined as an infinite series
 \begin{eqnarray} \nonumber
  & n-n+n-n+... =
n\cdot \sum\limits_{j =0}^{\infty}(-1)^n \; = n\cdot \lim\limits_{q \mapsto
1}\sum\limits_{j =0}^{\infty}{(-1)^n \over q^j}\; = n\cdot \lim\limits_{q \mapsto
1}{1\over 1+ q}={n\over 2}\;, \qquad
 \end{eqnarray}
so that $D$=$11$ model ($n$=$32$) is invariant under 16 $\kappa$--symmetries  and $32$
supersymmetries.

 In the Hamiltonian formalism, the above nilpotent matrix
 $\tilde{P}\!\!\!/^{\alpha\beta}$ can be used to extract
the (infinitely reducible) generator of the $\kappa$--symmetry,
$\tilde{P}\!\!\!/^{\alpha\beta} d_{\beta}$, from  the $n$ fermionic
primary constraints $d_\alpha =  \pi_\alpha + i
P\!\!\!/_{\alpha\beta}\theta^{\beta}$ which obey the algebra $\{
d_\alpha\, , \, d_\beta \}= 2iP\!\!\!/_{\alpha\beta}$,
\begin{eqnarray} \label{df=} d_\alpha &:=  \pi_\alpha + i
P\!\!\!/_{\alpha\beta}\theta^{\beta}\approx 0 \; , \qquad \{
d_\alpha\, , \, d_\beta \}= 2iP\!\!\!/_{\alpha\beta}\equiv 2i
\Gamma^m_{\alpha\beta}P_m \;
 \qquad
\end{eqnarray}
($\pi_\alpha:= {\partial L \over
\partial \dot{\theta}^{\alpha}  }$ and  $P_m:= {\partial L \over
\partial \dot{x}^{m} }$ are fermionic and bosonic momenta canonically conjugate to $\theta^\alpha (\tau)$ and
$x^m(\tau)$). Indeed, taking into account (\ref{PmPm=0}), one finds that $\{
\tilde{P}\!\!\!/^{\alpha\gamma}d_\gamma\, , \, d_\beta \}=2i \delta_\beta{}^\alpha
P^2\approx 0$, where $\approx$ denotes the weak equality in the sense of Dirac
\cite{Dirac}. On the other hand, the covariant extraction of the second class
constraint from the fermionic primary constraint (\ref{df=}) and, hence, the covariant
separation of the fermionic first and second class constraints, is not possible in the
frame of classical Brink-Schwarz formulation, {\it i.e.} without introducing additional
variables.

The infinite reducibility of the $\kappa$--symmetry and impossibility to separate
covariantly the fermionic first and the second class constraints are also
characteristics of the Green--Schwarz superstring. Many years these properties hampered
the way to the covariant superstring quantization. \footnote{See \cite{infGfG} for an
approach to quantization with infinitely many ghost for ghosts and its problems.}

\subsection*{3. `Pure spinor' BRST charge and its derivation in the
superparticle spinor moving frame formulation}


The  problem of covariant superstring quantization is now considered to be resolved by
the pure spinor formalism proposed by Berkovits in 2000, \cite{NB-pure}. It is based on
the {\it intrinsically complex} BRST charge
\begin{eqnarray}\label{QbrstB}
\mathbb{Q}^{B}= {\Lambda}^\alpha \; d_\alpha  \; ,
 \qquad
\end{eqnarray}
where $d_\alpha$ are the fermionic constraints (\ref{df=}) and ${\Lambda}^\alpha$ is
the complex {\it pure spinor} which obeys
\begin{eqnarray}\label{NB-pureSp}
    {\Lambda}\Gamma_a{\Lambda}=0 \; , \qquad
    {\Lambda}^\alpha\not= ({\Lambda}^\alpha)^*\; , \qquad \alpha=1,\ldots , n\;\;\; (n=32\;
    for \; D=11)
 \qquad
\end{eqnarray}
This constraint guarantees the nilpotency, $\{ \mathbb{Q}^{B},\mathbb{Q}^{B}\} =0$, of
the BRST charge (\ref{QbrstB}).

It is important that a solution of the constraint (\ref{NB-pureSp}) is provided by
\cite{IB07PLB}
\begin{eqnarray}\label{pureSp=}
 &   \widetilde{\Lambda}_\alpha = \tilde{\lambda}^+_p v_{\alpha
    p}^{\;-}\; , \qquad \tilde{\lambda}^+_p\tilde{\lambda}^+_p=0\; ,
    \quad \{v_{\alpha p}^{\;-}\} = {Spin(1,D-1) \over
[SO (1,1)\otimes Spin(D-2)] \, \subset\!\!\!\!\!\times
\mathbb{K}_{(D-2)} } = \mathbb{S}^{D-2} \; , \quad
\end{eqnarray}
where $\tilde{\lambda}^+_p$ is a complex $n/2$ component $SO(D-2)$ spinor {\it with
zero norm}, $\tilde{\lambda}^+_p\tilde{\lambda}^+_p=0$, and $v_{\alpha p}^{\;-}$ are
spinorial Lorentz harmonics \cite{BL98',Ghsds,GHT93} (see \cite{IB07PLB,IB07,BdAS2006}
for more references and discussion), a set of $n/2$ constrained $n$-component
$D$-dimensional bosonic spinors which, once the constraints are taken into account,
provide the homogeneous coordinates for the $D$ dimensional celestial sphere $S^{D-2}$
(see Eq. (\ref{v-inS11}) below).

The above equations are general, they are applicable, with $n=2, 4, 8, 16, 32$, for
$D=3,4,6,10,11$. However, some properties of spinor moving frame formalism are
dependent on $D$. In particular, in D=11 Eq. (\ref{pureSp=}) provides a particular
solution of the pure spinor constraint, while in $D$=10 it gives the general solution
\footnote{Indeed, in $D=11$ the generic null spinor $\Lambda_\alpha$ contains $23$
complex or $46$ real parameters \cite{NB-pure}, while eq. (\ref{pureSp=}) provides its
$39$ parametric solution. In D=10 the solution of Eq. (\ref{pureSp=}) carries {
16+8-2=22 } degrees of freedom, the same number as the generic pure spinor.}.

In \cite{IB07PLB,IB07} it was shown how the complex BRST charge (\ref{NB-pureSp}) for
$D=11$ ($n=32$) case can be obtained on the way of covariant BRST quantization of the
M0--brane, {\it i.e.} eleven dimensional massless superparticle, in its spinor moving
frame or twistor like Lorentz harmonic formulation \cite{BL98'} (described below, in
Sec. 4;  see
 \cite{BZ-str,BZ-strH}, \cite{BZ-p} and \cite{bst} for spinor moving frame
formulations of superstring, standard super-p-branes and super-Dp-branes).

 Namely, in \cite{IB07PLB,IB07} we
first constructed the Hamiltonian mechanics of this twistor-like formulation of the
$D=11$ superparticle  and, with the help of the spinorial Lorentz harmonics, separated
{\it covariantly} the first and the second class constraints (see \cite{BZ-strH} for an
analogous result for the Green-Schwarz superstring). Then we took into account the
second class constraints by introducing Dirac brackets \cite{Dirac}, and calculate the
Dirac brackets algebra of the first class constraints. Further, following the pragmatic
spirit of the Berkovit's approach \cite{NB-pure}, \cite{NBloops}, we take care of the
part of the first class constraints separately (partially, by imposing them as a
condition on the wavefunctions; one may also think on the gauge fixing at the classical
level) and left with a set of $16$ fermionic and $1$ bosonic first class constraints,
the generators of the fermionic $\kappa$--symmetry and its bosonic $b$--symmetry
superpartner, the Dirac brackets of which represent the $d=1$ , $n=16$ supersymmetry
algebra (the origin of $\kappa$--symmetry as worldline supersymmetry was found in
\cite{stv}). This set of constraints is described by the BRST charge
\cite{IB07PLB,IB07}
\begin{eqnarray}\label{Qsusy-Int}
\mathbb{Q}^{susy}= \lambda^+_q  D_q^{-} + i c^{++} \partial_{++} -
 \lambda^+_q\lambda^+_q {\partial\over \partial c^{++}}\; ,
 \qquad \{ D_p^{-} , D_q^{-}  \} = 2i \delta_{qp} \partial_{++} \; ,
 \qquad
\end{eqnarray}
including $16$ real  bosonic ghosts $\lambda^+_q$ and  one real
fermionic ghost $c^{++}$.

An analysis of  the cohomology of this BRST operator shows that it is trivial if the
norm $ \lambda^+_q\lambda^+_q$ of bosonic ghost $ \lambda^+_q$ is nonvanishing. In
other words, the nontrivial cohomology of $\mathbb{Q}^{susy}$ has support on
$\lambda^+_q\lambda^+_q=0$. For a {\it real} spinor $\lambda^+_q\lambda^+_q=0$ implies
$\lambda^+_q=0$. This produces a technical problem which is sorted out by means of a
regularization which consists in allowing $\lambda^+_q$ to be {\it complex},
$\lambda^+_q \mapsto \tilde{\lambda}^+_q\not= (\tilde{\lambda}^+_q)^*$. Furthermore,
this implies the reduction of the cohomology problem for the regularized BRST operator
$\mathbb{Q}^{susy}$ to the search for cohomology at vanishing bosonic ghost,
$\tilde{\lambda}^+_q=0$, for the following complex BRST charge
\begin{eqnarray}\label{tQsusy-Int}
\tilde{\mathbb{Q}}^{susy}= \tilde{\lambda}^+_q  \; D_q^{-} + i c^{++}
\partial_{++} \; , \qquad
 \tilde{\lambda}^+_q\tilde{\lambda}^+_q=0 \; ,
 \qquad \{ D_p^{-} , D_q^{-}  \} = 2i \delta_{qp} \partial_{++} \; .
 \qquad
\end{eqnarray}
Now, taking into account that $D^-_q$ represents the constraint
$d_q^-=v_q^{-\alpha}d_\alpha$, where $d_\alpha$ is the Brink--Schwarz fermionic
constraint (\ref{df=}), one finds that this non-Hermitian $\tilde{\mathbb{Q}}^{susy}$
operator is essentially (modulo additional $i c^{++}
\partial_{++}$ contribution) the Berkovits BRST operator (\ref{QbrstB}), but with
composite pure spinor (\ref{pureSp=}).

Thus \cite{IB07PLB,IB07} have shown (on the example of superparticle) the possible
origin of the intrinsic complexity of the Berkovits pure spinor BRST charge: it appears
on the stage of regularization in calculation of cohomology of the real BRST charge
(\ref{Qsusy-Int}).

Let us stress that of all the cohomologies of the Berkovits--like BRST charge
$\tilde{\mathbb{Q}}^{susy}$ (\ref{tQsusy-Int}) only the ones calculated (and remaining
nontrivial) at $\tilde{\lambda}_q=0$ describe the cohomology of the superparticle BRST
operator ${\mathbb{Q}}^{susy}$ \cite{IB07PLB,IB07}. The full cohomology of
$\tilde{\mathbb{Q}}^{susy}$ is clearly reacher and is related with spinorial
cohomologies of \cite{SpinCohom02}.

\subsection*{ 4. Spinor moving frame
of formulation of massless superparticle }

Since the constraint (\ref{PmPm=0}) is algebraic, it may be substituted into the action
(\ref{11DSSP-1st}), which gives $S^{\prime}_{M0}  =  \int_{W^1} \; P_{{m}}
\hat{\Pi}^{m}\vert_{P_{{m}} P^{{m}}=0}$. Thus, if the general solution of
(\ref{PmPm=0}) is known, one may substitute it for $P_m$ in (\ref{11DSSP-1st}) and
obtain a classically equivalent formulation of the Brink-Schwarz superparticle. It is
easy to solve the constraint (\ref{PmPm=0}) in a non-covariant manner: in a special
Lorentz frame a solution  with positive energy reads as, {\it e.g.},
$P^{\!\!\!^{0}}_{(a)} = {\rho\over 2} \; (1,\ldots , -1) = {\rho\over 2}  \;
(\delta_{(a)}^0 -\delta_{(a)}^{\#})$.  The solution in an arbitrary frame follows from
this by making a Lorentz transformation,
\begin{eqnarray}
\label{PmPm=01} P_m := U_m{}^{(a)} P^{\!\!\!^{0}}_{(a)} = {\rho\over 2} \; (u_{m}^{\;\;
0} - u_{m}^{\; \#}) \; , \qquad U_m^{\, (a)}:= (u_{m}^{\;\; 0} , u_{m}^{\;\; i} ,
u_{m}^{\;\#}) \in SO(1,D-1) \; .
\end{eqnarray}
Since  $P_{m}=P_{m}(\tau)$ is dynamical variable in the action (\ref{11DSSP-1st}),  the
same is true for the Lorentz group matrix, $U_m{}^{(a)}=U_m{}^{(a)}(\tau)= (u_{m}^{\;\;
0}(\tau) , u_{m}^{\;\; i}(\tau) , u_{m}^{\;\#}(\tau))$ in Eq. (\ref{PmPm=01}). Such
{\it moving frame variables} \cite{BZ-str} are called {\it Lorentz harmonics} (see
\cite{Sok,BL98',Ghsds}, \cite{IB07PLB,IB07} and refs. therein).

Substituting (\ref{PmPm=01}) for $P_{m}$ in (\ref{11DSSP-1st}), one arrives at the
action
\begin{eqnarray}\label{11DSSP(LH)}
& S  =  \int_{W^1} \; {1\over 2}  \rho^{++} u^{--}_{{m}}
\hat{\Pi}^{m} \; , \qquad u^{--}_{{m}} u^{--{m}}=0 \quad    \qquad
\end{eqnarray}
where the light--likeness of the vector $u^{--}_m=u^0_m - u^{\#}_m$,
 follows from the orthogonality and normalization of the
timelike $u_m^{0}$ and spacelike $u_m^{\# }$ vectors which, in their turn, follow  from
$U:= \{ u_m^{0}  , u_m^{\; i} , u_m^{\#}   \}=\{ {u_m^{++} + u_m^{--}\over 2} , u_m^{\;
i} , {u_m^{++} - u_m^{--}\over 2}  \} \in SO(1,10)$.

An important property of
 the action (\ref{11DSSP(LH)}) is that it {\it hides the twistor--like
action}
\begin{eqnarray}\label{11DSSP}
S:&=&   \int_{W^1} {1\over 2}\rho^{++}\, u_{m}^{--} \, \Pi^m =
\int_{W^1} {1\over 32}\rho^{++}\, v_{\alpha q}^{\; -} v_{\beta
q}^{\; -} \, \Pi^m \tilde{\Gamma}_m^{\alpha\beta}\; ,  \qquad
\end{eqnarray}
where, for $D=11$ case  $\alpha= 1,2, \ldots , 32$  ($n$ in general), $q=1, \ldots ,
16$ ($n/2$ in general ) and  $m=0,\ldots , 9, \# $,  ($\#=10$, $(D-1)$ in  general).
The first from of the action (\ref{11DSSP}) coincides with (\ref{11DSSP(LH)}); the
second form is twistor--like, {\it i.e.} it generalizes the Ferber--Schirafuji (FS)
action \cite{Ferber} to arbitrary $D$; the original $D=4$ FS action is reproduced  from
$n=4$ version of (\ref{11DSSP}) after writing the $D=4$ Majorana spinors in terms of
two Weyl ones. In $D=11$, instead of two--component unconstrained Weyl spinor in
\cite{Ferber}, the action of Eq. (\ref{11DSSP}) includes the set of $16$  bosonic
$32$--component Majorana spinors $v_{\alpha}{}^-_{q}$ which satisfy the following
kinematical constraints (see \cite{BZ-str,BZ-strH,BL98'}),
\begin{eqnarray}\label{vv=uG}
\left\{ \matrix{2 v_\alpha{}_{q}^{-} v_\beta{}_{q}^{-} &=&
 u_m^{--}{\Gamma}^m_{\alpha\beta}\;   & \qquad (a)\;
 ,  \cr  v_{q}^{-}\tilde{\Gamma}_m v_{p}^{-} &=&  \delta_{qp} \; u_m^{--}  \; & \qquad (b)\; ,
  \cr
v_\alpha{}_{q}^{-}C^{\alpha\beta}v_\beta{}_{p}^{-}&=& 0 \;  \qquad & \qquad (c)\; , }
\right. \;
   \quad u_m^{--}u^{m --}=0 \qquad (d)\;\; . \qquad
\end{eqnarray}

Although, in principle, one can study the dynamical system using just the kinematical
constraints (\ref{vv=uG}), for many problems, including the covariant quantization, it
is more convenient to treat the set of $16$ constrained $SO(1,10)$--spinors
$v_\alpha{}_{q}^{-}$ as part of the corresponding $Spin (1,10)$--valued matrix
describing the {\it spinor moving frame}, \begin{eqnarray} \label{harmVin}
V_\alpha^{(\beta)}= (v_\alpha{}_q^{-}\; , v_\alpha{}_{q}^{+})\; \in \; Spin(1,10) \quad
(\in \; Spin(1,D-1)\; in\; general ), \qquad \;
\end{eqnarray}
These {\it spinor moving frame} variables,
$v_\alpha{}_q^{-}\; , v_\alpha{}_{q}^{+}$, are also called  {\it spinor} {\it Lorentz
harmonics}.

\subsubsection*{4.1. Vector and spinor Lorentz harmonics. Spinor  moving
frame}

The relation between {\it vector} Lorentz harmonics $u_m^{\pm\pm}$, $u_m{}^i$
\cite{Sok},  which, in the $D$=11 case are elements of the $SO(1,10)$ Lorentz group
matrix
\begin{eqnarray}
\label{harmUin} && U_m^{(a)}= (u_m^{--}, u_m^{++}, u_m^{i})\;  \in \; SO(1,10) \quad
(\in \; SO(1,D-1)\; in\; general ), \qquad
\end{eqnarray}
and the {\it spinor}  harmonics \cite{Ghsds} or spinor moving frame variables
\cite{BZ-str,BZ-strH,BZ-p} $v^{\;\;\,\pm}_{\alpha q}$,  Eq. (\ref{harmVin}), are
defined by the Dirac matrices conservation
\begin{eqnarray}
\label{harmVdef} V \Gamma^{(a)} V^T = \Gamma^m U_m ^{(a)} \qquad (a) \; , \qquad V^T
\tilde{\Gamma}_m V = U_m^{(a)} \tilde{\Gamma}_{(a)} \qquad (b) \;   ,
\end{eqnarray}
and also by conservation of the charge conjugation matrix, if this exists,
\begin{eqnarray} \label{harmVdefC}
 VCV^T=C \quad, \quad V^TC^{-1}V=C^{-1}\; .
\end{eqnarray}
In this sense one says that the spinorial harmonics are `square roots' of the
associated vector harmonics.

Eqs. (\ref{harmVdef}) implies Eqs. (\ref{harmUin}), (\ref{harmVin}) modulo scaling
factor ($U\mapsto e^{2\gamma} U$, $V\mapsto e^{\gamma} V$). The fact that $U\in
SO(1,10)$ implies the following set of  constraints
\begin{eqnarray}
\label{harmUdef} U^T\eta U = \eta  \quad \Leftrightarrow \cases{ u_m^{--}u^{m--}=0 \; ,
\quad u_m^{++}u^{m++}=0 \; , \quad u_m^{\pm\pm}u^{m\, i}=0 \; , \cr u_m^{--}u^{m++}=2
\; , \qquad u_m^{i}u^{m\, j}=- \delta^{ij} }
\end{eqnarray}
or, equivalently,  $\delta_m^n= {1\over 2}u_m^{++}u^{n--} + {1\over 2}u_m^{--}u^{n++} -
u_m^{i}u^{n i}$ ($\Leftrightarrow \; U\eta U^T=\eta$).

The relations (\ref{harmVdef}), (\ref{harmVdefC}) reproduce the constraints
(\ref{vv=uG}). Indeed, using the Dirac matrices realization with diagonal  $\Gamma^0$
and $\Gamma^{\# }$, one may check that (\ref{vv=uG}a) coincides  the $(a)=(--)\equiv
(0)-(\# )$ component of Eq. (\ref{harmVdef}a) ; Eq. (\ref{vv=uG}b) comes from the upper
diagonal block of Eq. (\ref{harmVdef}b); finally, one of the diagonal blocks of
(\ref{harmVdefC}) gives rise to (\ref{vv=uG}c).

\subsubsection*{4.2. Gauge symmetries of the spinor moving frame action}

The action (\ref{11DSSP}) possesses  the {\it irreducible}
$\kappa$--symmetry\footnote{Let us stress that the possibility to reformulate the
$\kappa$--symmetry in the irreducible form is due to the presence of the constrained
bosonic spinor variables $v_\alpha{}_{q}^{-}$, spinorial harmonics  (see
\cite{BL98',BZ-str}).}
\begin{eqnarray}\label{kappa-irr}
\delta_\kappa x^m = i \delta_\kappa \theta^\alpha \Gamma^m_{\alpha\beta}\theta^\beta\;
, \qquad \delta_\kappa \theta^\alpha = \kappa^{+q} v_q^{-\alpha} \; , \qquad
\delta_\kappa v_\alpha{}^-_q =0 = \delta_\kappa u_m^{--}\; ; \qquad
\end{eqnarray}
as well as its superpartner called $b$--symmetry \cite{dA+L82}, which is the tangent
space copy of the worldvolume reparametrization symmetry, $\delta_b x^m = b^{++}
u^{--m}$ , $\delta_b \theta^\alpha = 0$, $\delta_b v_\alpha{}^-_q =0 = \delta_b
u_m^{--}$, and a scaling $GL(1,\mathbb{R})$ symmetry
\begin{eqnarray}\label{SO(1,1)}
\rho^{++} \mapsto e^{2\alpha} \rho^{++}\; , \qquad u_m^{--} \mapsto e^{-2\alpha}
u_m^{--}\; , \qquad v_{\alpha q}{}^- \mapsto e^{- \alpha} v_{\alpha q}{}^- \; ,  \qquad
\end{eqnarray}
with the wait determined by the sign indices $^{++}$, $^{--}$ and
$^{-}$, which we prefer to identify as $SO$(1,1) subgroup of
$SO$(1,D-1). The action (\ref{11DSSP}) is also invariant under the
$Spin(9)$ symmetry acting on the $q$ index of the constrained
bosonic spinor variable  $v_{\alpha q}{}^-$,
\begin{eqnarray}\label{SO(9)}
v_{\alpha q}{}^{\!\!\! -} \mapsto  v_{\alpha p}{}^{\!\!\! -} S_{pq}\; , \qquad S_{pq}
\in Spin(9)\; \quad \Leftrightarrow  \quad \cases{ S^TS=\mathbb{I}_{16\times 16} \; ,
\cr S\gamma^I S^T = \gamma^J U^{JI} \; , \quad U^TU= \mathbb{I}_{9\times 9} }\; ,
\qquad
\end{eqnarray}
Notice that the nine dimensional charge conjugation matrix is symmetric and can be
identified with the Kroneker delta symbol, $ \delta_{qp}\; $, so that the contraction
$v_{\alpha q}{}^{\!\!\! -}v_{\beta q}{}^{\!\!\! -}$, entering the action, is $Spin(9)$
invariant.

 Finally, when $v_{\alpha p}{}^{\!\!\! -}$ are considered as Lorentz harmonics
the fact of absence of the other $16\times 32$ block $v_{\alpha
p}{}^{\!\!\! +}$ of the spinor moving frame matrix (\ref{harmVin})
can be formulated as the statement of $K_9$ symmetry,
\begin{eqnarray}
\label{K9-def} \delta v_{\alpha q}^{\; -}=0\; , \qquad \delta
v_{\alpha q}^{\; +}=
 k^{++ i} \gamma^i{}_{qp}\,v_{\alpha p}^{\; -}\; , \qquad i=1,\ldots , 9 \; . \qquad
\end{eqnarray}
The $[SO(1,1)\otimes SO(D-2)]\subset\!\!\!\!\!\!\times K_{D-2}$ is
the Borel subgroup of $SO(1,D-1)$ so that \cite{Ghsds} the coset
$SO(1,D-1)\over [SO(1,1)\otimes SO(D-2)]\subset\!\!\!\!\times
K_{D-2}$ is compact; moreover it is isomorphic to the sphere
$S^{D-2}$ which can be identified as celestial sphere of the
$D$--dimensional observer \cite{Ghsds}.

Thus, using the  $Spin(9)$, $SO(1,1)$ and $K_9$ symmetry, Eqs.
(\ref{SO(9)}), (\ref{SO(1,1)}) and (\ref{K9-def}), as an
identification relation, the spinor harmonics $v_{\alpha q}{}^{\!\!
-}$, explicitly present in the action (\ref{11DSSP}), can be
identified as homogeneous coordinates of the celestial sphere $S^9$
($S^ {D-2}$) of the eleven-dimensional (D-dimensional) observer
\cite{Ghsds,GHT93},
\begin{eqnarray}
\label{v-inS11} {} \{v_{\alpha q}^{\;-}\} = {Spin(1,10) \over [Spin (1,1)\otimes
Spin(9)] \, \subset \!\!\!\!\!\!\times {\mathbb{K}_9} } = \mathbb{S}^{9}  \quad  &
\left(= {Spin(1,D-1) \over [Spin (1,1)\otimes Spin(D-2)] \, \subset \!\!\!\!\times
{\mathbb{K}_{D-2}} } = \mathbb{S}^{D-2}\right)  \; . \quad
\end{eqnarray}

\subsubsection*{4.3. On O(16) gauge symmetry of the M0--brane  action} \label{O(16)}

However, when the action (\ref{11DSSP}) with the variable $v_{\alpha q}{}^-$ subject
only to the constraints (\ref{vv=uG}) is considered (we call them $\tilde{v}_{\alpha
q}{}^-$ to distinguish from the harmonics $v_{\alpha q}{}^-$), one notices that neither
constraints nor the action involve the $d=9$ gamma matrices; all the contractions are
made with $16\times 16$ Kroneker $\delta_{qp}$ only. This implies  that the $D=11$
action (\ref{11DSSP}), when considered as constructed from $16$ 32-component spinors
$\tilde{v}_{\alpha q}^{\; -}$ restricted by (\ref{vv=uG}) only,
\begin{eqnarray}\label{11DSSP-16}
S &=& \int_{W^1} {1\over 32}\rho^{++}\, \tilde{v}_{\alpha q}^{\; -} \tilde{v}_{\beta
q}^{\; -} \, \Pi^m \tilde{\Gamma}_m^{\alpha\beta}\; , \qquad  \cases{ 2
\tilde{v}_\alpha{}_{q}^{-} \tilde{v}_\beta{}_{q}^{-} =
 {1\over 16} \tilde{v}_{p^\prime}^{-}\tilde{\Gamma}_m \tilde{v}_{p^\prime}^{-} {\Gamma}^m_{\alpha\beta}\;
 , \quad (a)\; \cr  \tilde{v}_{q}^{-}\tilde{\Gamma}_m \tilde{v}_{p}^{-} =   \delta_{qp} \;
{1\over 16} \tilde{v}_{p^\prime}^{-}\tilde{\Gamma}_m \tilde{v}_{p^\prime}^{-} \; ,
\quad (b)
 \cr
\tilde{v}_\alpha{}_{q}^{-}C^{\alpha\beta}\tilde{v}_\beta{}_{q}^{-}=0\; ,
   \qquad {} \qquad {}\quad  (c) \; } \; ,  \qquad
\end{eqnarray}
 actually {\it possesses the local
$SO(16)$ symmetry} acting on the $q=1,\ldots , 16$ indices of
$\tilde{v}_{\alpha q}^{\; -}$,
\begin{eqnarray}\label{SO(16)}
\tilde{v}_{\alpha q}{}^- \mapsto  \tilde{v}_{\alpha p}{}^- O_{pq}\; , \qquad O_{pq} \in
O(16)\quad \Leftrightarrow  \quad  O^TO=\mathbb{I}_{16\times 16} \;  . \qquad
\end{eqnarray}
The relation between spinorial harmonic ${v}_{\alpha q}{}^- $, which
transforms under $Spin(9)$ symmetry, and the above
$\tilde{v}_{\alpha p}{}^- $, carrying the $SO(16)$ index $p$ is
given by \cite{BdAS2006,IB07}
\begin{eqnarray}\label{tv-=v-L}
\tilde{v}_{\alpha p}{}^- =  {v}_{\alpha q}{}^- L_{qp}\; , \qquad L_{qp} \in O(16)\quad
\Leftrightarrow  \quad  L^TL=\mathbb{I}_{16\times 16} \;  , \qquad
\end{eqnarray}
where $L_{qp}$ is an arbitrary orthogonal $16\times 16$ matrix.

Eq. (\ref{tv-=v-L}) provides the general solution of the constraints
(\ref{vv=uG}a-d) \cite{BdAS2006,IB07}. As
  $\tilde{v}_{\alpha p}{}^- \tilde{v}_{\beta p}{}^- =
{v}_{\alpha q}{}^- {v}_{\beta q}{}^-$, substituting (\ref{tv-=v-L}) for
$\tilde{v}_{\alpha p}{}^-$ in (\ref{11DSSP-16}), one observes the cancelation  of the
contributions of the matrix $L_{qp}$. This shows the $O(16)$ symmetry of the action
(\ref{11DSSP-16}) with variable restricted only by the constraints presented there
explicitly. Furthermore \cite{BdAS2006,IB07}, this (seemingly fictitious) $SO(16)$
symmetry of the M0--brane, which we have observed studying different versions of
(defferent treatment of the variable in) its twistor--like formulation, reappears
inevitably in the quantization of the physical degrees of freedom.

\subsubsection*{4.4. Supertwistor representation of the M0-brane action}

The spinor moving frame superparticle action (\ref{11DSSP}) can be
written in the following equivalent form \cite{BdAS2006}:
\begin{eqnarray}\label{S=rhoTw}
S &=&  \int_{W^1} (\lambda_{\alpha q}\,  d {\mu}^{\alpha}_{q}-
 d\lambda_{\alpha q}\; {\mu}^{\alpha}_{q} - i d\eta_{q}\,\eta_{q}) \; ,
\end{eqnarray}
were the sixteen 32-component spinors $\lambda_{\alpha q}$ are taken
to be proportional to the spinor harmonics $v_{\alpha q}^{\;-}$ in
product with an arbitrary $SO(16)$ valued matrix,
\begin{eqnarray}
\label{Tw=H} \lambda_{\alpha q}:= \sqrt{\rho^{++}} v_{\alpha p}^{\;
-}L_{pq}\;  ,  \qquad LL^T=I_{16\times 16}\; . \quad
\end{eqnarray}
Hence these $16$ bosonic spinorial variables  obey the constraints
(see (\ref{vv=uG}) or (\ref{11DSSP-16}a-c))
\begin{eqnarray}\label{ll=pG}
2 \lambda_\alpha{}_{q} \lambda_\beta{}_{q} = p_m {\Gamma}^m_{\alpha\beta}\; , \qquad
\lambda_{q} \tilde{\Gamma}_m \lambda_{p} = \delta_{qp} \; p_m  \; ,  \qquad
C^{\alpha\beta} \lambda_\alpha{}_{q}
   \lambda_\beta{}_{p}=0\; ,  \qquad p_mp^m=0\; ,
\end{eqnarray}
with a light-like vector $p_m=\rho^{++}u_m^{--}$ which can be
identified as a massless particle momentum. On account of
$\rho^{++}$ in (\ref{Tw=H}), and due to Eq. (\ref{v-inS11}), the
$\{\lambda_\alpha{}_{q}\}$ parametrize the $\mathbb{R}_+ \times
\mathbb{S}^{9}$ manifold (all the degrees of freedom in the SO(16)
matrix $L_{pq}$ are pure gauge); furthermore due to (\ref{vv=uG}),
this is identified as the space parametrized by the light--like
momentum $p_m\,, \; p^2=0$,
\begin{eqnarray}
\label{l-inS11}
 {} \{\lambda_{\alpha p}\} = \; \mathbb{R}_+\; \times \; \mathbb{S}^{9} \; = \;  \{p_m\; :\; p_np^n=0\}\; . \qquad
\end{eqnarray}
The variables $ {\mu}^{\alpha}_{q}$, $\eta_{q}$ in (\ref{S=rhoTw}) are related to the
superspace coordinates by the following generalization of the Penrose incidence
relation,
\begin{eqnarray}\label{Tw=}
{\mu}^{\alpha}_{q}:= {1\over 32} x^m\tilde{\Gamma}_m^{\alpha\beta} \lambda_{\beta q} -
{i\over 2} \theta^\alpha \, \theta^\beta \lambda_{\beta q}  \; , \qquad \eta_{q}:=
\theta^\beta \lambda_{\beta q} \; .
\end{eqnarray}
Together with $\lambda_{\alpha q}\,$, the  ${\mu}^{\alpha}_{q}$ and ${\eta}_{q}$ in (
\ref{Tw=}) define a set of sixteen constrained $OSp(1|64)$ supertwistors (see
\cite{BdAS2006}), $
 \Upsilon_{\Sigma \,q} := ( \lambda_{\alpha q}\; , \;
{\mu}^{\alpha}_{q}\; , \;  {\eta}_{q})$. The action (\ref{S=rhoTw}) can be written as
$S=\int_{W^1}  d{\Upsilon}_{\Sigma\, q}\Omega^{\Sigma\Pi}\Upsilon_{\Pi\,q}$ where
$\Omega^{\Sigma\Pi}= -(-)^{(\Sigma +1)(\Pi +1)}\Omega^{\Pi\Sigma}$ is the
orthosymplectic $OSp(1|64)$ invariant tensor (including the symplectic
$\Omega^{\alpha\beta}= - \Omega^{\beta\alpha}$ invariant of $Sp(32)$).

\subsection*{ 5. Supertwistor covariant quantization of M0--brane and
hidden symmetries of D=11 supergravity}

The supertwistor quantization of D=11 superparticle has been performed in
\cite{BdAS2006}; there it was firstly motivated that, in the purely bosonic limit, the
wavefunction is just an arbitrary function on the $\mathbb{R}_+ \otimes \mathbb{S}^9$
space, which allows for identification with the space of light--like momenta, although
appears as parametrized by its ''squere root'', Eqs. (\ref{ll=pG}), provided by the
highly constrained coordinates $\lambda_{\alpha q}$, Eqs. (\ref{l-inS11}),
\begin{eqnarray}\label{Phi=S9R} \Phi \vert_{\theta_q =0} = \Phi_0(\mathbb{R}_+
\otimes \mathbb{S}^9)\; , \qquad \{ (v_{\alpha q}^-\, , \; \rho^{++}) \} = \mathbb{R}_+
\otimes \mathbb{S}^9 = \{ (p_{\underline{m}}\; : \; p^2:=
p_{\underline{m}}p^{\underline{m}}=0 )\}\;  . \qquad
\end{eqnarray}
Then, beyond the purely bosonic limit, the pure supertwistor form (\ref{S=rhoTw}) of
the superparticle action contains the set of 16 free fermionic fields $\eta_q$ which,
upon quantization, become the $Cl^{16}$ Clifford algebra valued variables,
\begin{eqnarray}\label{hThhTh=}
{} & \{ \hat{\eta}_{q}\, , \,
 \hat{\eta}_{p} \} = {1\over 2}  \delta_{qp} \; , \qquad q=1,2,\ldots , 16 . \qquad
\end{eqnarray}
This $O(16)$ covariant Clifford algebra $\mathrm{C}\ell^{16}$ has a
finite dimensional representation by $256\otimes 256$ sixteen
dimensional gamma matrices $\hat{\eta}_{q} = \, {1\over 2 }\,
({\Gamma}_{q})_{{\cal A}}{}^{{\cal B}}$ (${\cal A}\, , \, {{\cal B}}
= 1, \ldots , 256\,$,  $\;q=1,\ldots , 16 $).
 The choice of this representation in the M0-brane quantization
implies that the  wavefunction is to be the $\mathbf{256}$ Majorana
spinor representation of $SO(16)$ (see \cite{BdAS2006}),
\begin{eqnarray}\label{Phi(256)=b+f}
\Phi_{{\cal A}} \, := & \left( \matrix{\Phi_A \cr \Psi^B } \right)\;
= \left(\matrix{ \left( \matrix{ h_{IJ}  \cr A_{IJK}} \right) \cr {}
\cr \sqrt{2} \Psi_{Iq} } \right)\; , \qquad \cases{ h_{IJ} =
h_{(IJ)}\; , \quad h_{II}=0 \; , \cr A_{IJK}=A_{[IJK]}\; , \cr
\Psi_{Iq}\gamma^I_{qp}=0 \;  } \qquad
\end{eqnarray}
(see \cite{IB07}), and this describes the linearized $D$=11
supergravity multiplet with $h_{IJ} = h_{(IJ)}, A_{IJK}=A_{[IJK]}$
and $\Psi_{Iq}$ restricted by  $h_{II}=0=\Psi_{Iq}\gamma^I_{qp}$
(see \cite{Green+99}).

This indicates the $SO(16)$ symmetry of the linearized $D$=11
supergravity  multiplet and suggests \cite{BdAS2006} the origin of
the $SO(16)$ symmetry of (uncompactified) D=11 supergravity observed
by Nicolai in \cite{Nicolai87}. Notice that our spinor moving frame
formulation (\ref{11DSSP}) makes this the $SO(16)$ symmetry manifest
already at the classical level.

Furthermore, as it is well-known, $E_8$ exceptional group Lie algebra can be written in
terms of the generators of $SO(16)$ and $128$  bosonic generators carrying the Majorana
spinor ($\mathbf{128}$) representation of $SO(16)$ \cite{gsw},
\begin{eqnarray}
\label{E8:SO} E_8\; : & \;  [J_{qp}\, , \, J_{q^\prime p^\prime }\,
]= 4 \delta_{_[q^\prime{}\, ^[q}J_{p^] \, p^\prime{}_]} \; , \quad
[J_{qp}\, , \, Q_A]= {1\over 2}\sigma_{pq}{}_{AB} Q_B \; ,  \quad [
Q_A\, , \, Q_B]= \sigma_{pq}{}_{AB} J_{pq} \; . \quad
\end{eqnarray}
This makes tempting to speculate \cite{IB07} on that the $E_8$
symmetry might be characteristic of the $D=11$ supergravity itself
rather than of its reduction to $d=3$ only. A check of whether this
is the case  is an interesting subject for future study. See
\cite{IB07} for more discussion.

\subsection*{6. Outlook }

Probably the most important conclusion  of the study of the M0--brane covariant
quantization in \cite{IB07PLB,IB07} is that the twistor-like Lorentz harmonic approach
\cite{BZ-str,BdAS2006} is able to produce a simple and practical BRST charge. This
suggests a similar investigation of the $D=10$ Green--Schwarz superstring case. The
natural guess is that such a quantization of {\it e.g.} type IIB superstring should
produce (after some stages of reduction/simplification)  the Berkovits BRST charge
\begin{eqnarray}\label{QIIB}
 \mathbb{Q}^B_{IIB}=\int \Lambda^{\alpha 1}d_{\alpha} + \int \Lambda^{\alpha
2}d^2_{\alpha}\; , \qquad \Lambda^{\alpha 1}\sigma^a_{\alpha\beta}\Lambda^{\beta 1}= 0=
\Lambda^{\alpha 1}\sigma^a_{\alpha\beta}\Lambda^{\beta 1}\; ,
 \end{eqnarray}
but with composite pure spinors $\Lambda^{\alpha 1}$ and $\Lambda^{\alpha 2}$ given by
\begin{eqnarray}\label{pureSp12=}
    \widetilde{\Lambda}^{\alpha 1} = \tilde{\lambda}^+_p v^{-\alpha}_p\; , \qquad
 \widetilde{\Lambda}^{\alpha 2} = \tilde{\lambda}^-_p v^{+\alpha}_p\; ,
\qquad    \tilde{\lambda}^+_p\tilde{\lambda}^+_p=0=
\tilde{\lambda}^+_p\tilde{\lambda}^+_p\; . \qquad
\end{eqnarray}
Here, the $\tilde{\lambda}^{\pm}_p$ are two complex eight component $SO(8)$ spinors and
the stringy harmonics $v^{\mp\alpha}_p$ are the homogeneous coordinates of the
non--compact $16$--dimensional coset
\begin{eqnarray}\label{harmV=IIB}
& \{ V_{(\beta)}{}^{\alpha} \} = \{ ( v^{+\alpha}_p \; , \; v^{-\alpha}_p )\}  =
{Spin(1,9) \over SO(1,1)\otimes SO(8) } \; ,
\end{eqnarray}
characteristic for the spinor moving frame formulation of the
(super)string \cite{BZ-str,BZ-strH}.

\medskip

 {\bf Acknowledgments.} The author thanks  Jos\'e de Azc\'arraga for collaboration on the early
stages of this work and Dima Sorokin for useful discussions. This work has been
partially supported by research grants from the Ministerio de Educaci\'on y Ciencia
(FIS2005-02761) and EU FEDER funds, the Generalitat Valenciana, the Ukrainian State
Fund for Fundamental Research (N383), the INTAS (2006-7928) and by the EU
MRTN-CT-2004-005104  network.

\medskip

{\bf Notice added}. When the present work was finished, the author became aware of the
work \cite{Duff85} in which the possible hidden $E_8\times SO(16)$ symmetry of D=11
supergravity was conjectured for the first time.

\end{document}